\documentclass[aoas,nameyear,rotating,seceqn,dvips]{arximspdf}
\usepackage{dcolumn}
\usepackage{graphicx}

\doi{10.1214/10-AOAS342}
\volume{4}
\issue{4}
\pubyear{2010}
\firstpage{1892}
\lastpage{1912}

\makeatletter
\renewcommand{\le}{\leq}

\newtheorem{proposition}{Proposition}

\newcommand{\eqref}[1]{(\ref{#1})}
\newcolumntype{d}[1]{D{.}{.}{#1}}
\makeatother

\begin{document}
\begin{frontmatter}

\title{A nonlinear mixed effects directional model for the estimation
of the rotation axes
of the human~ankle}
\runtitle{Mixed effects directional model}

\begin{aug}
\author[A]{\fnms{Mohammed} \snm{Haddou}\ead[label=e1]{mohammed.haddou.1@ulaval.ca}},
\author[A]{\fnms{Louis-Paul} \snm{Rivest}\corref{}\thanksref
{t1}\ead[label=e2]{Louis-Paul.Rivest@mat.ulaval.ca}}
\and\\
\author[C]{\fnms{Michael}~\snm{Pierrynowski}\ead[label=e3]{pierryn@mcmaster.ca}}
\runauthor{M. Haddou, L.-P. Rivest and M. Pierrynowski}
\affiliation{Universit\'e Laval, Universit\'e Laval and McMaster University}
\thankstext{t1}{Supported in part by the Natural Sciences and
Engineering Research
Council of Canada and of the Canada Research Chair in Statistical
Sampling and Data Analysis.}
\address[A]{M. Haddou \\
L.-P. Rivest \\
D\'epartement de math\'ematiques \\
\quad et de statistique \\
Universit\'e Laval \\
Qu\'ebec (Qu\'ebec), G1V 0A6\\
Canada \\
\printead{e1} \\
\phantom{\textsc{E-mail: }}\printead*{e2}} 
\address[C]{M. Pierrynowsi \\
School of Rehabilitation Science \\ Institute of Applied Health
Sciences \\ McMaster University \\ 1400 Main Street West \\ Hamilton, Ontario,
L8S 1C7 \\ Canada\\ \printead{e3}}
\end{aug}

\received{\smonth{9} \syear{2009}}
\revised{\smonth{1} \syear{2010}}

%
\begin{abstract}
This paper suggests a nonlinear mixed effects model for data points in
$\mathit{SO}(3)$,
the set of $3 \times3$ rotation matrices, collected according to
a repeated measure design. Each sample individual contributes
a sequence of rotation matrices giving the relative orientations of the
right foot with respect to the right lower leg as its ankle moves.
The random effects are the five angles characterizing the
orientation of the two rotation axes of a subject's right ankle.
The fixed parameters are the average value of these angles
and their variances within the population. The algorithms to fit
nonlinear mixed effects models presented
in Pinheiro and Bates (\citeyear{Pienheiro2000}) are adapted to the new directional model.
The estimation of the random
effects are of interest since they give predictions of
the rotation axes of an individual ankle. The performance of these
algorithms is
investigated in a Monte Carlo study. The analysis of two data sets is presented.
In the biomechanical literature, there is no consensus
on an in vivo method to estimate the two rotation axes of the ankle.
The new model is promising. The estimates obtained from a
sample of volunteers are shown to be in agreement with the clinically
accepted results of Inman (\citeyear{Inman1976}), obtained by manipulating cadavers.
The repeated measure
directional model presented in this paper is developed for a particular
application.
The approach is, however, general and might be applied to other models
provided that the random directional effects are clustered around their
mean values.
\end{abstract}

%

\end{frontmatter}

\section{Introduction}
\label{section-1}

The human ankle joint complex has been modeled as a two fixed axis mechanism.
It is the
primary joint involved in the motion of the rearfoot with respect to
the lower leg. The
characterization of walking disorders associated with cerebral
palsy, clubfoot or flatfoot deformities uses altered external
moments (torques) about the two rotation axes of the ankle. An
accurate and reliable determination of the orientation of these two
axes is important to successfully evaluate and treat patients with
these conditions. There is no consensus, in the biomechanical
literature, on a noninvasive method for estimating the location and orientation
of these ankle axes in a live individual.

\begin{figure}

\includegraphics{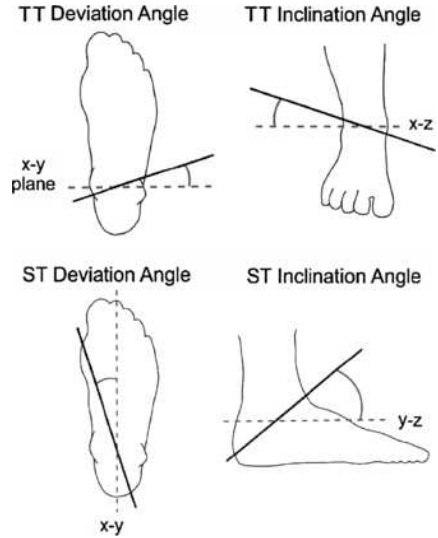}

\caption{The deviation and inclination angles of the
tibiotarsal (TT) and subtalar (ST) human ankle rotation axes.}\label{Axes}
\end{figure}

The two rotation axes of the ankle can be recorded in an RFU
coordinate system where the $x$-axis points Right, the $y$-axis goes
Forward and the $z$-axis goes Up. Anatomically, plantarflexion--dorsiflexion
occurs about the tibiotarsal, or $\mathit{tt}$
axis, which is attached to the lower leg, while the subtalar, or $\mathit{st}$
axis, attached to the calcaneus, is used for the supination-pronation
motion of the foot. These two axes are
presented in Figure~\ref{Axes}. Their orientations are determined by four anatomical
angles (\textit{ttinc}, \textit{ttdev}, \textit{stinc}, \textit{stdev})
giving the inclinations and the deviations of these two axes referenced
to the RFU coordinate system.

Using an average generic orientation for each axis results in
substantial errors [Lewis et al. (\citeyear{Lewis2009})] because of an important
between subject variation in axis locations as characterized in
Section~\ref{sec7.1} below. Several in vivo estimation methods have been
proposed in biomechanical journals [see, for instance, van den Bogert, Smith
and Nigg (\citeyear{Bogert1994}) and Lewis et al. (\citeyear{Lewis2009})], but none was
completely successful at estimating the angles in Figure~\ref{Axes}. The poor
numerical results obtained by Lewis, Sommer and Piazza (\citeyear{Lewis2006}) led them to
question the validity of the two-axis model for the ankle.

The in vivo estimation of the orientation of the two axes of the ankle
is a statistical problem. The data set for one individual is a sequence
of $3 \times3$ rotation matrices giving the relative orientation of
the foot of the subject with
respect to its lower leg as its ankle moves up and down, right and
left, as much as possible. Rivest, Baillargeon and Pierrynowski (\citeyear{Rivest2008})
and van den Bogert, Smith
and Nigg (\citeyear{Bogert1994}) provide more details about data collection.

Following van den Bogert, Smith
and Nigg (\citeyear{Bogert1994}), Rivest, Baillargeon and Pierrynowski (\citeyear{Rivest2008}) developed a
statistical model
for analyzing the rotation data collected on a single ankle.
Its parameters are the four angles defined in Figure~\ref{Axes} and a fifth one for
the relative position of the two rotation axes. This model fits well;
the residual standard deviations reported in Rivest, Baillargeon and Pierrynowski (\citeyear{Rivest2008}) are
around one degree.
For a subject whose ankle has an average range of motion, the
estimates are, however, not repeatable. Two data sets
collected on the same ankle in similar conditions can give different
estimated angles. This occurs because the likelihood
function does not have a clear maximum; it has a plateau and some
angles cannot be estimated independently of the others. Indeed,
Rivest, Baillargeon and Pierrynowski (\citeyear{Rivest2008}) demonstrate that only three parameters can be
reliably estimated in a subject with an average ankle range of
motion. Considering the small range of the residual angles,
a failure of the two-axis model is an unlikely cause for
the poor repeatability of the results.

This paper suggests methods to improve the numerical stability of the
estimates derived
from the ankle model. A penalized
likelihood is proposed for estimating the parameters. The penalty is
obtained by assuming a prior multivariate normal distribution for
the five angle parameters. When the mean and the variance covariance
matrix of the prior
distribution are not known, one is confronted with a nonlinear
mixed effects directional model whose parameters can be estimated using ankle's
data collected on a sample of volunteers. Two numerical algorithms to
fit this model are
proposed. Their performances are evaluated in a Monte Carlo experiment;
two data sets are then analyzed using the new nonlinear mixed effects
directional model.

Nonlinear mixed effects vary with the
parametrization of the random effects. Thus, the first step of the
analysis is to parameterize the model of Rivest, Baillargeon and Pierrynowski (\citeyear{Rivest2008}) in
terms of the angles presented in Figure~\ref{Axes} and to derive inference
techniques using this parametrization. These procedures are then
generalized to a Bayesian setting obtained by multiplying the
likelihood by the prior distribution of the model parameters. The
algorithms of Lindstrom and Bates (\citeyear{Lindstrom1990}) [see also Pinheiro and
Bates (\citeyear{Pienheiro2000})] for fitting nonlinear mixed effects models are then adapted
to the new directional model.

Nonlinear mixed effects and Bayesian models with
concentrated prior distributions could potentially be used in many
problems of
directional statistics. These techniques could, for instance, be
applied to the spherical
regression model considered by Kim (\citeyear{Kim1991}) and Bingham, Chang and
Richards (\citeyear{Bingham1992}) and to the directional one way ANOVA model of
Rancourt, Rivest and Asselin (\citeyear{Rancourt2000}) to characterize the between
subject variability of the mean rotation.

This new statistical methodology has important applications in
biomechanics. By fitting a nonlinear mixed effects directional
model to data collected on a sample of volunteers, one is able to
estimate the mean values and the between subject variances of the
five anatomical angles in the population. These estimates are found
to be close to the clinically accepted results obtained by Inman (\citeyear{Inman1976})
 who used direct measurements from cadaveric feet. This
provides an empirical validation of the two-axis model of van den Bogert, Smith
and Nigg (\citeyear{Bogert1994}) for the ankle. The new model also allows the
analysis of the rotation data collected by Pierrynowski et al.
(\citeyear{Pierrynowski2003}) which compared the orientations of the subtalar axis of two
groups of individuals classified according to their lower
extremity injuries. Finally, the penalized predictions associated to the
mixed effect model are shown to
be more stable than the estimates obtained by maximizing the
standard unpenalized likelihood for one subject.

\section{Parameterization of unit vectors and of rotation matrices
using anatomical angles}
\label{section-2}

Let $\mathbf{A}_1$ and $\mathbf{B}_2$ be $3 \times1$ unit
vectors giving the tibiotarsal and the subtalar rotation axes in a
coordinate system defined according to the RFU convention. These
unit vectors are first expressed in terms of the anatomical angles.
Then the Cardan angle decomposition for a $3 \times3$ rotation
matrix is briefly reviewed. This section uses the $\arctan$
function with two arguments, such that $\arctan(a,b)$ is the angle
whose sine and cosine are given by $a/\sqrt{a^2+b^2}$ and
$b/\sqrt{a^2+b^2}$, respectively.

First consider the tibiotarsal axis,
$\mathbf{A}_1=(A_{11},A_{21},A_{31}){^\top}$, where ``${^\top
}$'' denotes a
matrix transpose. Formally, the anatomical angles are defined as
$\mathit{ttinc}=t_1=-\arctan(A_{31},A_{11})$ and
$\mathit{ttdev}=t_2=\arctan(A_{21},A_{11})$. Without loss of generality, we
assume that the first coordinate of $\mathbf{A}_1$ is positive.
A general expression for unit vectors in the half unit sphere is
%
\begin{equation}
\label{a1}
\mathbf{A}_1=\frac1 {D_t}
\pmatrix{
\cos(t_1) \cr \cos(t_1) \tan(t_2) \cr -\sin(t_1)
}
,  \qquad t_1,t_2 \in[-\pi/2,\pi/2),
\end{equation}
where $D_t=\sqrt{1+\cos^2(t_1)\tan^2(t_2)}$. In a similar manner, one
can parameterize the subtalar axis in terms of the anatomical angles
$\mathit{stinc}=s_1=\arctan(B_{32},B_{22})$ and $\mathit{stdev}=s_2=-\arctan
(B_{12},B_{22})$ as follows,
%
\begin{equation}
\label{b2}
\mathbf{B}_2=\frac1 {D_s}
\pmatrix{
-\cos(s_1) \tan(s_2) \cr \cos(s_1) \cr \sin(s_1)
}
, \qquad  s_1,s_2 \in[-\pi/2,\pi/2),
\end{equation}
where $D_s=\sqrt{1+\cos^2(s_1)\tan^2(s_2)}$.

The set $\mathit{SO}(3)$ of $3 \times3$ rotation matrices is a
three-dimensional
manifold whose properties are reviewed in McCarthy (\citeyear{McCarthy1990}), Chirikjian
and Kyatkin (\citeyear{Chirikjian2001})
and Le\'on, Mass\'e and Rivest (\citeyear{Le2006}). This paper uses the Cardan angle
parametrization
with the $X-Z-Y$ convention. It expresses an element of $\mathit
{SO}(3)$ as a function of the Cardan angles $\alpha\in
[-\pi,\pi)$,
$\gamma\in[-\pi/2,\pi/2)$ and $\phi\in[-\pi,\pi)$ as
%
\begin{eqnarray}
\label{cardan}\hspace*{18pt}
\mathbf{R} &=&
\pmatrix{ 1 & 0 & 0 \cr
0 & \cos\alpha& -\sin\alpha \cr
0 &\sin\alpha& \cos\alpha
}
 \pmatrix{\cos\gamma& -\sin\gamma& 0 \cr
\sin\gamma& \cos\gamma& 0 \cr
0 & 0 & 1
}
\pmatrix{ \cos\phi& 0 & \sin\phi \cr
0 & 1 & 0 \cr
-\sin\phi& 0 & \cos\phi
}
\nonumber
\\[-8pt]
\\[-8pt]\hspace*{18pt} &=&
 \pmatrix{ \cos\gamma\cos\phi& -\sin\gamma& \cos\gamma
\sin\phi\cr
\cdots& \cos\alpha\cos\gamma &
\cdots\cr
\cdots& \sin\alpha\cos\gamma & \cdots
}
,\nonumber \!\!\!\!\!\!\!
\end{eqnarray}
where $\cdots$ stands for complex trigonometric expressions that are
not used in the sequel. We also write
$\mathbf{R}=\mathbf{R}(\alpha,x)\times
\mathbf{R}(\gamma,z)\times\mathbf{R}(\phi,y)$, where the
arguments of $\mathbf{R}(\cdot,\cdot)$ are the angle and the
axis of the rotation respectively.

The model presented in the next section uses rotation matrices,
$\mathbf{A}(t_1,t_2)$ and $\mathbf{B}(s_1,s_2),$ whose first
and second columns are respectively equal to $\mathbf{A}_1$ and
$\mathbf{B}_2$. These matrices are given by
\begin{eqnarray*}
\mathbf{A}(t_1,t_2)&=&\mathbf{R}(t_1,y)\mathbf
{R}[\arctan\{\cos(t_1)\tan(t_2),1\},z]\\
&=&\frac1 {D_t} \pmatrix{
\cos(t_1) & -\cos^2(t_1)\tan(t_2)& \sin(t_1)D_t\cr
\cos(t_1) \tan(t_2)& 1&0 \cr -\sin(t_1) & \sin(t_1)\cos(t_1)\tan(t_2)
& \cos(t_1)D_t
}
\end{eqnarray*}
and
\begin{eqnarray*}
\mathbf{B}(s_1,s_2)
&=&\mathbf{R}(s_1,x)\mathbf{R}[\arctan\{\cos(s_1)\tan
(s_2),1\},z]\\
&=&\frac1 {D_s} \pmatrix{
1& -\cos(s_1) \tan(s_2)&0\cr
\cos^2(s_1) \tan(s_2)& \cos(s_1)& -\sin(s_1)D_s \cr
\sin(s_1)\cos(s_1) \tan(s_2)& \sin(s_1) &
\cos(s_1)D_s
}
,
\end{eqnarray*}
where $D_t$ and $D_s$ are defined in (\ref{a1}) and (\ref{b2}),
respectively.

\section{The model for estimating the rotation axes of a single ankle}
\label{section-3}

This section expresses the model of Rivest, Baillargeon and Pierrynowski (\citeyear{Rivest2008}) for the
estimation of the anatomical angles for a single subject in terms of
the rotation matrices $\mathbf{A}(t_1,t_2)$ and
$\mathbf{B}(s_1,s_2)$. The data set is a sequence of time
ordered $3 \times3$ rotation matrices $\{\mathbf{R}_i\dvtx
i=1,\ldots,n\}$. The model for $\mathbf{R}_i$ involves the four
anatomical angles, rotation angles $\{\alpha_i \dvtx  i=1,\ldots,n\}$
and $\{\phi_i \dvtx  i=1,\ldots,n\}$ in $[-\pi,\pi)$ about the two
rotation axes and the angle $\gamma_0 \in(-\pi/2,\pi/2)$, related
to the relative position of the two axes. The predicted value
$\bolds{\Psi}_i$ for $\mathbf{R}_i$ is given by
%
\begin{equation}
\label{psi} \bolds{\Psi}_i
=\mathbf{A}(t_1,t_2)\mathbf{R}(\alpha_i,x)\mathbf
{R}(\gamma_0,z)
\mathbf{R}(\phi_i,y)\mathbf{B}(s_1,s_2){^\top}.
\end{equation}

The errors are assumed to have a symmetric Fisher--von Mises matrix
distribution with density $f(\mathbf{E})=\exp\{\kappa
\operatorname{tr}(\mathbf{E})\}/c_\kappa$, $\mathbf{E}\in\mathit{SO}(3)$,
where $c_\kappa$ is a normalizing constant; see Mardia and Jupp
(\citeyear{Mardia2000}). If the parameter $\kappa$ is assumed to be large so that the
error rotations are clustered around the identity matrix
$\mathbf{I}_3$, one has
%
\begin{equation}\label{error}
\mathbf{E}=\mathbf{I}_3+ \pmatrix{ 0 &
-\varepsilon_3 & \varepsilon_2\cr \varepsilon_3 & 0 &-\varepsilon_1
\cr- \varepsilon_2 & \varepsilon_1 &0
}
 + O_p \biggl(\frac1 \kappa \biggr),
\end{equation}
where the entries $\varepsilon_1$, $\varepsilon_2$, $\varepsilon_3$
of the skew-symmetric matrix have independent $\mathcal{N} \{
0,1/(2\kappa)\}$
distributions; $1/(2\kappa)$ is called the residual variance. The
model postulates that
$\mathbf{R}_i=\bolds{\Psi}_i\mathbf{E}_i$, for
$i=1,\ldots,n$. The likelihood is
$L[t_1,t_2,s_1,s_2,\gamma_0,\break \kappa,\{\alpha_i \},\{\phi_i\}]=\prod_i
f(\bolds{\Psi}_i{^\top}\mathbf{R}_i)$. Rivest, Baillargeon and Pierrynowski (\citeyear{Rivest2008}) show
that the angles $\{\alpha_i\}$ and $\{\phi_i\}$ can be profiled out
of the likelihood. Indeed,
%
\begin{eqnarray}
\label{prof}
\qquad L[t_1,t_2,s_1,s_2,\gamma_0,\kappa,\{\alpha_i
\},\{\phi_i\}] &\le& L_p(t_1,t_2,s_1,s_2,\gamma_0,\kappa) \nonumber
\\[-8pt]
\\[-8pt] &=& \frac1
{c_\kappa^n}   \exp \Biggl[ \kappa\sum_{i=1}^n
\{2\cos(\theta_i^z-\gamma_0)+1\} \Biggr],\nonumber
\end{eqnarray}
where $\theta_i^z=-\arcsin( \mathbf{A}_1{^\top} \mathbf{R}_i
\mathbf{B}_2)$ is the $Z$-Cardan angle of
$\mathbf{A} (t_1,t_2){^\top} \mathbf{R}_i$ $\mathbf{B}
(s_1,s_2)$
in the $X-Z-Y$ convention; see (\ref{cardan}).

Several methods are available to maximize (\ref{prof}). However, for
the implementation of the mixed effects directional model, a closed
form expression for the score vector for
$\bolds{\beta}=(t_1,t_2,s_1,s_2,\gamma_0){^\top}$ is needed.
This is
derived now. Observe that $\cos(\theta)=1-2\sin^2(\theta/2)$, where
$\theta/2$ can be assumed to lie in the interval $[-\pi/2,\pi/2)$.
The log profile likelihood is equal to
\[
\log L_p(t_1,t_2,s_1,s_2,\gamma_0,\kappa)=-4\kappa\sum_{i=1}^n \sin
^2
\biggl(\frac{\theta_{i}^z - \gamma_0}2  \biggr) -n \log c_\kappa+3n
\kappa.
\]
The score for $\gamma_0$ is easily evaluated, viz.
\[
\frac{\partial}{\partial\gamma_0} \log
L_p(t_1,t_2,s_1,s_2,\gamma_0,\kappa) = 4\kappa\sum_{i=1}^n \sin
\biggl(\frac{\theta_{i}^z - \gamma_0}2  \biggr)\cos
\biggl(\frac{\theta_{i}^z - \gamma_0}2  \biggr).
\]
The score for the four remaining parameters involve the following
partial derivatives that are evaluated using elementary methods.
The property that the partial derivatives of a unit vector and
the unit vector itself are orthogonal was used to get the following results:
\begin{eqnarray*}
\frac{\partial}{\partial t_1}   \mathbf{A}_1
&=&-\frac{\sin
(t_1) \tan(t_2)}{D_t^2}   \mathbf{A}_2
-\frac{1}{D_t}   \mathbf{A}_3, \\
\frac{\partial}{\partial t_2}   \mathbf{A}_1
&=& \frac{\cos
(t_1)\{1+ \tan^2(t_2)\}}{D_t^2}   \mathbf{A}_2, \\
\frac{\partial}{\partial s_1}   \mathbf{B}_2
&=&\frac{\sin
(s_1) \tan(s_2)}{D_s^2}   \mathbf{B}_1
+\frac{1}{D_s}   \mathbf{B}_3, \\
\frac{\partial}{\partial s_2}   \mathbf{B}_2
&=& - \frac{\cos
s_1\{1+ \tan^2(s_2)\}}{D_s^2}   \mathbf{B}_1.
\end{eqnarray*}

Since $\theta_i^z=-\arcsin( \mathbf{A}_1{^\top} \mathbf{R}_i
\mathbf{B}_2)$, the score for $t_1$ is given by
\begin{eqnarray*}
\frac{\partial}{\partial t_1} \log
L_p(t_1,t_2,s_1,s_2,\gamma_0,\kappa) &=& -4\kappa\sum_{i=1}^n \sin
\biggl(\frac{\theta_{i}^z - \gamma_0}2  \biggr)\cos
\biggl(\frac{\theta_{i}^z - \gamma_0}2  \biggr) \\
&&\hphantom{-4\kappa\sum_{i=1}^n}
{}\times\frac{-1}{\sqrt{1-(
\mathbf{A}_1{^\top} \mathbf{R}_i \mathbf{B}_2)^2}}
\frac{\partial}{\partial t_1}   \mathbf{A}_1{^\top}
\mathbf{R}_i
\mathbf{B}_2.
\end{eqnarray*}
This can be evaluated using the previous expressions for the partial
derivatives. Repeating this for the other anatomical angles leads
to
%
\begin{eqnarray}\label{score}
\frac{\partial}{\partial\bolds{\beta}}   \log
L_p(\beta,\kappa)
&=& -4\kappa\sum_{i=1}^n \sin \biggl(\frac
{\theta_{i}^z
- \gamma_0}2  \biggr) \cos \biggl(\frac{\theta_{i}^z - \gamma_0}2
 \biggr) \frac{\partial}{\partial\bolds{\beta}} (\theta_{i}^z
- \gamma_0)\nonumber
\\[-8pt]
\\[-8pt]
&=&-4\kappa\sum_{i=1}^n \sin \biggl(\frac{\theta_{i}^z - \gamma
_0}2
\biggr)\mathbf{X}_i,\nonumber
\end{eqnarray}
where
%
\begin{eqnarray}\label{xi}
\mathbf{X}_i &=& -\frac{\cos\{(\theta_{i}^z - \gamma_0)/2\}} {
\sqrt{1-( \mathbf{A}_1{^\top} \mathbf{R}_i \mathbf{B}_2)^2}}
\nonumber
\\[-8pt]
\\[-8pt]  &&\phantom{-}{}\times\pmatrix{\displaystyle -\frac{\sin(t_1) \tan
(t_2)}{ D_t^2}   \mathbf{A}_2{^\top}\mathbf{R}_i
\mathbf{B}_2 -\frac{1}{ D_t
}   \mathbf{A}_3{^\top}\mathbf{R}_i \mathbf{B}_2\cr\displaystyle
\frac{\cos
(t_1)\{1+ \tan^2(t_2)\}}{ D_t^2}
\mathbf{A}_2{^\top}\mathbf{R}_i \mathbf{B}_2 \cr\displaystyle
\frac{\sin(s_1) \tan(s_2)}{ D_s^2}
  \mathbf{A}_1{^\top} \mathbf{R}_i\mathbf{B}_1
+\frac{1}{ D_s }   \mathbf{A}_1{^\top}
\mathbf{R}_i\mathbf{B}_3\cr\displaystyle
- \frac{\cos
(s_1)\{1+ \tan^2(s_2)\}}{ D_s^2}   \mathbf
{A}_1{^\top}
\mathbf{R}_i\mathbf{B}_1\cr
\sqrt{1-( \mathbf{A}_1{^\top}
\mathbf{R}_i \mathbf{B}_2)^2}
}
.\nonumber
\end{eqnarray}
Evaluated at $\bolds{\beta}+\delta(\bolds{\beta})$, where
$\delta(\bolds{\beta})$ is a $5 \times1$ vector with entries
close to 0, the score vector (\ref{score}) is
\begin{eqnarray*}
&&-4\kappa\sum_{i=1}^n  \biggl\{\sin \biggl(\frac{\theta_{i}^z -
\gamma_0}2
 \biggr)\mathbf{X}_i + \frac1 2
\mathbf{X}_i\mathbf{X}_i{^\top}\delta
(\bolds{\beta}) \biggr\}
\\
&&\qquad{}+ O(\|\delta(\bolds{\beta})\|^2)
+O\bigl(\max[|\sin\{(\theta_{i}^z - \gamma_0)/2\}|  \| \delta
(\bolds{\beta})\|]\bigr).
\end{eqnarray*}
One can consider that the last two terms are negligible since the
residuals $(\theta_{i}^z - \gamma_0)$ are small. This is standard in
the large $\kappa$ asymptotics used to approximate the sampling
distributions of estimators in a directional model: both
$\hat{\bolds{\beta}}-\bolds{\beta}$ and the errors~$\varepsilon_j$ in (\ref{error}) are assumed to be
$O(1/\sqrt{\kappa})$; see, for instance, Rivest and Chang (\citeyear{Rivest2006}).
Equating this to 0 yields a simple updating formula. Given its
current value~$\bolds{\beta}$, the updated value is
$\bolds{\beta}+\delta(\bolds{\beta})$, where
\[
\delta(\bolds{\beta}) = - \Biggl(\sum_{i=1}^n
\mathbf{X}_i\mathbf{X}_i{^\top} \Biggr)^{-1} \sum_{i=1}^n
 \biggl\{2\sin
 \biggl(\frac{\theta_{i}^z - \gamma_0}2  \biggr)\mathbf{X}_i
 \biggr\}.
\]

This calculation can be carried out by regressing the residual vector
$[2\sin
\{(\theta_{i}^z - \gamma_0)/2\}]$ on the explanatory variables
$\{\mathbf{X}_i\}$. Theorem 1 of Rivest, Baillargeon and Pierrynowski (\citeyear{Rivest2008}) holds and,
as $\kappa$ goes to $\infty$, the maximum likelihood estimator~$\hat
{\bolds{\beta}}$ is approximately normally distributed. Once
the model is fitted, the residual variance $1/(2\kappa)$ can be
estimated using the sum of the squared residuals,
\[
\frac1{2\hat\kappa}=\frac{1}{n}   \sum_{i=1}^n 4\sin^2 \{(\hat
\theta_{i}^z
- \hat\gamma_0)/2 \},
\]
where the residual angle $\hat\theta_{i}^z=-\arcsin( \widehat
{\mathbf{A}}_1{^\top}
\mathbf{R}_i \widehat{\mathbf{B}}_2)$ is the $Z$-Cardan angle of
$\widehat{\mathbf{A}}{^\top}
\mathbf{R}_i \widehat{\mathbf{B}}$ in the $X-Z-Y$ convention.
Using Grood and Suntay (\citeyear{Grood1983}) clinical interpretation, rotations of
angle $\{\hat\theta_{i}^z\}$ occur about a floating axis that is
orthogonal to both the $\mathit{tt}$ and the $\mathit{st}$ axes. Plots of these angles
appear in Figure~2 of Rivest, Baillargeon and Pierrynowski (\citeyear{Rivest2008}); their residual standard
deviation, $\sqrt{1/2 \hat\kappa}$, is about one degree. The
distribution of the residuals $\{[2\sin\{(\theta_{i}^z - \gamma
_0)/2\}]\}$ is usually approximately normal; the normality assumption
in (\ref{error}) is not violated for most of the data sets
investigated. Thus, the proposed model fits well to the ankle data.

Many individuals have a limited ankle range of motion; the domains for
angles $\{\alpha_i\}$ and $\{\phi_i\}$ in (\ref{psi}) are therefore limited.
This makes the estimates of the anatomical angles
numerically unstable. There can be important differences between
the estimates calculated on two data sets collected in succession on
the same ankle; see Rivest, Baillargeon and Pierrynowski (\citeyear{Rivest2008}). Thus, individual
measurements do not allow the estimation of a complete set of
anatomical angles. This suggests to borrow strength from other
individuals and to consider a Bayesian model whose prior
distribution penalizes extreme parameter values.

\section{A Bayesian ankle model}
\label{section-4}

Assume, for now, that the residual variance $1/(2\kappa)$ is known
and that the $5 \times1$ vector of anatomical angles
$\bolds{\beta}$ is random and has a
$\mathcal{N}_5 (\bolds{\beta}_0,\bolds{\Sigma}_0)$, where
$\bolds{\beta}_0$ is the average vector of anatomical angles
within the population and the $5 \times5$ variance covariance
matrix $\bolds{\Sigma}_0$ characterizes the variability of the
anatomical angles within the population; both are assumed to be
known. Elements of $\bolds{\beta}_0$ and
$\bolds{\Sigma}_0$ could be set equal to the values of Inman (\citeyear{Inman1976}),
who studied the  variability of these angles. We assume that
$\bolds{\Sigma}_0$ is $O(1/(2\kappa))$, thus, there is a fixed
$5 \times5$ upper triangular matrix $\bolds{\Delta}_0$ such
that
$\bolds{\Sigma}_0=\bolds{\Delta}_0^{-1}(\bolds
{\Delta}_0^{-\top})/(2\kappa)$,
where $\bolds{\Delta}_0^{-\top}$ is the inverse of
$\bolds{\Delta}_0^{\top}$. This section presents an algorithm
to derive the mode of the posterior
distribution of $\bolds{\beta}$ and suggests an approximation
for its posterior distribution.

The posterior distribution of $\bolds{\beta}$ is proportional to
\begin{eqnarray*}
&&\exp \Biggl[-\kappa \Biggl\{\sum_{i=1}^{n}4
\sin^2 \biggl(\frac{\theta_{i}^z -\gamma_{0}}2 \biggr)
+ (\bolds{\beta}-
\bolds{\beta}_0){^\top}\Delta_0{^\top}\Delta_0(\bolds
{\beta}-
\bolds{\beta}_0) \Biggr\} \Biggr] \\
&&\qquad = \exp\{-\kappa
\mathit{SSE}(\bolds{\beta})\}.
\end{eqnarray*}
The posterior mode $\widehat{\bolds{\beta}}$ is the value of
$\bolds{\beta}$ that minimizes $\mathit{SSE}$. It can be evaluated by
adapting the regression algorithm of Section~\ref{section-3} to the Bayesian
framework. The vector of partial derivatives of $\mathit{SSE}$ with respect
to $\bolds{\beta}$ is easily derived,
%
\begin{equation}\label{score2}
\frac{\partial}{\partial\bolds{\beta}}   \mathit{SSE}(\bolds
{\beta}) =4
\sum_{i=1}^{n}
\sin \biggl(\frac{\theta_{i}^z -
\gamma_{0}}2  \biggr)\mathbf{X}_{i} +
2\Delta_0{^\top}\Delta_0(\bolds{\beta}-
\bolds{\beta}_0),
\end{equation}
where the vector of partial derivatives $\mathbf{X}_{i}$ is
defined by (\ref{xi}).

Proceeding as in Section~\ref{section-3}, one constructs an algorithm for
minimizing $\mathit{SSE}$. The current value $\bolds{\beta}$ is updated
to $\bolds{\beta}+\delta(\bolds{\beta})$, where
\begin{eqnarray*}
&&\delta(\bolds{\beta}) = - \Biggl(\sum_{i=1}^n
\mathbf{X}_{i}\mathbf{X}_{i}{^\top} +\Delta_0{^\top}\Delta_0
 \Biggr)^{-1} \Biggl[\sum_{i=1}^{n} \biggl \{2\sin
\biggl(\frac{\theta_{i}^z - \gamma_0}2  \biggr)\mathbf{X}_{i}
\biggr\}
  \\
  &&\hspace*{182pt}{}+   \Delta_0{^\top}\Delta_0(\bolds{\beta}-
\bolds{\beta}_0)  \Biggr].
\end{eqnarray*}
An alternative expression for the updated value is
\begin{eqnarray*}
\bolds{\beta}+ \delta(\bolds{\beta}) &=& \Biggl (\sum_{i=1}^n
\mathbf{X}_{i}\mathbf{X}_{i}{^\top} +\Delta_0{^\top}\Delta_0
 \Biggr)^{-1}
 \\
&&{}\times \Biggl[
 \sum_{i=1}^{n}
 \biggl\{-2
 \sin
\biggl(\frac{\theta_{i}^z - \gamma_0}{2}\biggr)
+\mathbf{X}_{i}
{^\top}\bolds{\beta}
  \biggr\}
\mathbf{X}_{i}+ \Delta_0{^\top}\Delta_0
\bolds{\beta}_0
\Biggr].
\end{eqnarray*}
An approximation to the posterior distribution of the anatomical
angles is given next.

\begin{proposition}\label{prop1}  As $\kappa\to\infty$,
the posterior distribution for $\beta$ satisfies
\[
\sqrt{2\kappa}   (\bolds{\beta} - \hat{\bolds{\beta
}})\sim
\mathcal{N}_5\Biggl \{0,  \Biggl(\sum_{i=1}^n \hat{\mathbf
{X}}_{i}\hat
{\mathbf{X}}_{i}{^\top} + \Delta_0{^\top} \Delta_0
\Biggr)^{-1} \Biggr\},
\]
where $\hat{\mathbf{X}}_{i}$ denotes the $5 \times1$ vector of
partial derivatives defined by \eqref{xi} and evaluated at $\hat
{\bolds{\beta}} $.
\end{proposition}

The posterior density of $\delta= \sqrt{2\kappa}  (\bolds
{\beta} -
\hat{\bolds{\beta}})$ is proportional to $\exp\{-\kappa
\mathit{SSE}(\hat{\bolds{\beta}} + \delta/\sqrt{2\kappa})\}$. The result
is proved by taking a second-order Taylor series expansion around
$\delta=0$. The first order derivatives are null and the variance
covariance matrix of Proposition~\ref{prop1} is obtained by dropping the
$o\{1/(2\kappa)\}$ terms in the matrix of second-order derivatives.

We now study some frequentist properties of $\hat
{\bolds{\beta}}$. Let $\bolds{\beta}^{(t)}$ be the true
values of the anatomical angles for the individual under
consideration. Thus, $\bolds{\beta}^{(t)}$ is a realization of
the $\mathcal{N}_5 (\bolds{\beta}_0,\bolds{\Sigma}_0)$ prior
distribution, such that
$\|\bolds{\beta}^{(t)}-\bolds{\beta}_0\|$ is
$O(1/\sqrt{2\kappa})$. The difference $ \hat
{\bolds{\beta}}-\bolds{\beta}^{(t)}$ is
$O(1/\sqrt{2\kappa})$. The leading term of this difference consists
of a linear combination of individual experimental errors and of
the penalty associated to the prior distribution. To get a closed
form expression, one can proceed as in Appendix B of Rivest, Baillargeon and Pierrynowski (\citeyear{Rivest2008}).
It suffices to carry out a first-order Taylor series
expansion of (\ref{score2}) in terms of the difference
$\delta(\bolds{\beta})={\bolds{\beta}}-\bolds
{\beta}^{(t)}$
and of the experimental errors. This yields
%
\begin{eqnarray}
\label{score3}
&&\frac{\partial}{\partial\bolds{\beta}}
\mathit{SSE}\bigl(\bolds{\beta}^{(t)}+\delta(\bolds{\beta})\bigr)
\nonumber
\\
&&\qquad = 2 \Biggl\{ \sum_{i=1}^{n} \epsilon_i \mathbf{X}_{i}^{(t)}+
\sum_{i=1}^{n}
\mathbf{X}_{i}^{(t)}\mathbf{X}_{i}^{(t)T}\delta(\bolds
{\beta})
+
\Delta_0{^\top}\Delta_0\bigl(\delta(\bolds{\beta})+\bolds
{\beta}^{(t)}-
\bolds{\beta}_0\bigr)
 \Biggr\} \\
 &&\qquad\quad {}+ O(1/\kappa),\nonumber
\end{eqnarray}
where $\epsilon_i$ is a $\mathcal{N}(0,1/(2\kappa))$ random variable that
depends on the error matrix $\mathbf{E}_i$, and
$\mathbf{X}_{i}^{(t)}$ denotes the vector of partial derivatives
$\mathbf{X}_{i}$ evaluated at the true value
$\bolds{\beta}^{(t)}$, with $R_i$ set equal to $\Psi_i$ in
(\ref{xi}). Now $\hat{\bolds{\beta}}$ corresponds to the value
of $\delta(\bolds{\beta})$ for which (\ref{score3}) is null,
thus,
\begin{eqnarray*}
\hat{\bolds{\beta}} &=& \bolds{\beta}^{(t)} - \Biggl(\sum_{i=1}^{n}
\mathbf{X}_{i}^{(t)}\mathbf{X}_{i}^{(t)T} +
\bolds{\Delta}_0{^\top}\bolds{\Delta}_0  \Biggr)^{-1}
 \\
 &&\hphantom{\bolds{\beta}^{(t)} -\,}{}\times
 \Biggl\{ \sum_{i=1}^{n} \epsilon_i \mathbf{X}_{i}^{(t)} +
\bolds{\Delta}_0{^\top}\bolds{\Delta}_0\bigl(\bolds
{\beta}^{(t)}-
\bolds{\beta}_0\bigr) \Biggr\} + O(1/\kappa)\\
&= & \Biggl(\sum_{i=1}^{n}
\mathbf{X}_{i}^{(t)}\mathbf{X}_{i}^{(t)T} +
\bolds{\Delta}_0{^\top}\bolds{\Delta}_0  \Biggr)^{-1}
 \\
 &&{}\times\Biggl\{ \sum_{i=1}^{n}
\bigl(\mathbf{X}_{i}^{(t)T}\bolds{\beta}^{(t)}-\epsilon_i\bigr)
\mathbf{X}_{i}^{(t)} +
\bolds{\Delta}_0{^\top}\bolds{\Delta}_0
\bolds{\beta}_0 \Biggr\}+ O(1/\kappa).
\end{eqnarray*}
This expansion provides an approximation for the prediction error of
$\hat{\bolds{\beta}}$ as a predictor of
$\bolds{\beta}^{(t)}$. The posterior variance covariance matrix
of $\bolds{\beta}^{(t)}$ given in Proposition~\ref{prop1} is an estimate
of the variance covariance matrix of the approximate prediction error.

\section{A mixed model for the simultaneous estimation of several sets
of anatomical
angles}
\label{section-5}

We now have $M$ subjects and $n_i$, $1 \le i \le M$, observed
rotation matrices on each subject. The data set consists of the $3
\times3$ rotation matrices $\{\mathbf{R}_{ij}\dvtx  i=1,\ldots,M ;
j=1,\ldots, n_i\}$. Let
$\bolds{\beta}_i=(t_{1i},t_{2i},s_{1i},s_{2i},\gamma
_{0i}){^\top}$
be the anatomical angles for the $i$th ankle. As in Section~\ref
{section-4}, the
angles $\bolds{\beta_i}$ are assumed to be random deviates with
a five-dimensional normal distribution,
$\mathcal{N}_5 (\bolds{\beta}_0,\bolds{\Sigma}_0)$. The fixed
parameters $\kappa$, $\bolds{\beta}_0$ and
$\bolds{\Sigma}_0$ are assumed to be unknown.

In a mixed effects model, the fixed regression parameters and the
variance components are
estimated using a marginal likelihood. To estimate
$\bolds{\beta}_0$ and $\bolds{\Sigma}_0$, we construct such
a likelihood using the profile likelihood defined in (\ref{prof}),
rather than the full likelihood for the ankle model. This is
acceptable since this profile likelihood is also a likelihood,
constructed by assuming that the angles $\theta_{ij}^z-\gamma_{0i}$
defined in (\ref{prof}) have a centered angular von Mises
distribution with shape parameter $2\kappa$. Indeed, since $\kappa$
is large, the distribution of
$2\sin\{(\theta_{ij}^z-\gamma_{0i})/2\}$ is approximately
$\mathcal{N}\{0,1/(2\kappa)\}$. Using this approximation in the
evaluation of
the marginal likelihood $L_{1p}$ gives
%
\begin{eqnarray}
\label{margi}
&&\prod_{i=1}^{M}  \biggl(\frac{\kappa}{\pi}
\biggr)^{(n_i+5)/2}|\bolds{\Delta}_0|
 \nonumber
\\[-8pt]
\\[-8pt]
&&\qquad{}\times \int_{\mathbb{R}^5}   \exp \Biggl\{ -\kappa\sum_{j=1}^{n_i}4
\sin^2 \biggl(\frac{\theta_{ij}^z -\gamma_{0i}}2 \biggr)
-\kappa\| \Delta_0 (\bolds{\beta}_i-
\bolds{\beta}_0)\|^2  \Biggr\}  \,{\mathrm{d}} \bolds{\beta}_i.\nonumber
\end{eqnarray}
Because of a highly nonlinear integrand, the above expression cannot be
evaluated explicitly. To find the maximum value
numerically, we adapt the algorithm of Lindstrom and Bates (\citeyear{Lindstrom1990}) to
the directional ankle model.

For each $i$, and for fixed
$(\bolds{\beta}_0,\bolds{\Delta}_0)$, the integrand of
(\ref{margi}) is maximized using the method presented in Section~\ref
{section-3}.
Let $ \hat{\bolds{\beta}}_i$ be the maximum value for the
$i$th sample. Using the first order Taylor series expansion derived
in Section~\ref{section-3} leads to the following approximation:
\[
2 \sin \biggl(\frac{\theta_{ij}^z-\gamma_{0i}}2 \biggr) \approx2
\sin \biggl(\frac{\hat\theta_{ij}^z-\hat\gamma_{0i}}2 \biggr)
+\mathbf{X}_{ij}{^\top}({\bolds{\beta}}_i- \hat
{\bolds{\beta}}_i),
\]
where the angles $\hat\theta_{ij}^z$ and $\hat\gamma_{0i}$ and the
$5 \times1$ vector of partial derivatives $\mathbf{X}_{ij}$ are
evaluated at $\hat{\bolds{\beta}}_i$. Using this
approximation, and changing variables
$\bolds{\beta}_i-\bolds{\beta}_0=\mathbf{z}$ in the
integral, (\ref{margi}) becomes
\begin{eqnarray*}
L_{1p} &\approx&\prod_{i=1}^{M} \biggl (\frac{\kappa}{\pi}
\biggr)^{(n_i+5)/2}
|\bolds{\Delta}_0|  \\
&&\hphantom{\prod_{i=1}^{M}}
{}\times\int_{\mathbb{R}^5}   \exp \Biggl\{ -\kappa
\sum_{j=1}^{n_i}(\mathbf{y}_{ij}-\mathbf{X}_{ij}{^\top}{\mathbf
{z}-\mathbf{X}_{ij}{^\top}\bolds{\beta}}_0)^2
-\kappa\| \bolds{\Delta}_0 \mathbf{z}\|^2  \Biggr\} \,{\mathrm{d}} \mathbf{z},\\
\\ &= &\prod_{i=1}^{M} \frac{(\kappa/\pi)^{n_i/2}}
{|\mathbf{I}+\mathbf{X}_{i}\bolds{\Delta
}_0^{-1}\bolds{\Delta}_0^{-\top}\mathbf{X}_{i}{^\top}|^{1/2}}
     \\
     &&\hphantom{\prod_{i=1}^{M} }{}\times \exp\{ -\kappa
(\mathbf{y}_{i}-\mathbf{X}_{i}{\bolds{\beta}}_0){^\top}
(\mathbf{I}+\mathbf{X}_{i}\bolds{\Delta}_0^{-1}\bolds{\Delta
}_0^{-\top}\mathbf{X}_{i}{^\top})^{-1}
(\mathbf{y}_{i}-\mathbf{X}_{i}{\bolds{\beta}}_0)
\},
\end{eqnarray*}
where $\mathbf{X}_{i}$ is the $n_i\times5$ matrix of partial
derivatives for the $i$th unit, and $\mathbf{y}_i$ is the
\mbox{$n_i\times1$} vector whose $j$ entry is given by
$\mathbf{X}_{ij}{^\top}\hat{\bolds{\beta}}_i- 2 \sin\{
(\hat
\theta_{ij}^z-\hat\gamma_{0i})/2 \}$. This evaluation of the
integral with respect to $\mathbf{z}$ follows the argument
presented in Pinheiro and Bates (\citeyear{Pienheiro2000}), Section~7.2. It involves the
five-dimensional normal density with mean vector
$(\mathbf{X}_i{^\top}\mathbf{X}_i + \bolds{\Delta
}_0{^\top}
\bolds{\Delta}_0)^{-1}\mathbf{X}_i{^\top}
(\mathbf{y}_{i}-\mathbf{X}_{i}{\bolds{\beta}}_0)$ and
variance covariance matrix equal to
$(\mathbf{X}_i{^\top}\mathbf{X}_i + \bolds{\Delta
}_0{^\top}
\bolds{\Delta}_0)^{-1}/(2\kappa)$. Thus, $L_{1p}$ is
approximately equal to the likelihood function for the following
linear mixed effects model:
%
\begin{equation} \label{mixed}
\mathbf{y} = \pmatrix{\mathbf{X}_1 \cr \vdots\cr \mathbf{X}_M
}
 \bolds{\beta}_0
+ \pmatrix{\mathbf{X}_1& \mathbf{0} & \cdots
&\mathbf{0} \cr
\mathbf{0}& \mathbf{X}_2 & \cdots& \mathbf{0}\cr
\cdots& \cdots& \cdots&\cdots\cr
\mathbf{0} & \mathbf{0} & \cdots& \mathbf{X}_M
}
 \mathbf{b} + \bolds{\epsilon},
\end{equation}
where $\mathbf{y}$ is the $(\sum n_i) \times1$ vector of the
dependent variable, $\bolds{\beta}_0$ is the mean vector of the
anatomical angles in the population, and $\mathbf{b}$ is the
$5M\times1$ vector of the individual random effects
$\mathbf{b}_i=\bolds{\beta}_i-\bolds{\beta}_0$ that
are assumed to be independent random vectors with a
$\mathcal{N}_5 (0,\bolds{\Sigma}_0)$ distribution. Finally,
$\bolds{\epsilon}$ is the $(\sum n_i) \times1$ vector of the
experimental errors of the directional model containing independent
$\mathcal{N}(0,1/(2\kappa))$ random deviates. The second step of the algorithm
estimates $\bolds{\beta}_0$, $1/(2\kappa)$  and
$\bolds{\Sigma}_0$ by fitting the linear mixed effects model
(\ref{mixed}). This gives updated values for the parameters of the
prior distribution that are used to get a new set of penalized
estimates $\{\hat{\bolds{\beta}}_i\}$; these new estimates are
used to get a new approximation to (\ref{margi}) and to update the
marginal parameter values by fitting (\ref{mixed}). This two-step
algorithm typically converges after a few iterations. It is called
the PLME algorithm since it uses a Penalized least squares and an
algorithm for fitting Linear Mixed Effects models.
Lindstrom and Bates (\citeyear{Lindstrom1990}) showed that
the first step can be bypassed by using $\hat
{\bolds{\beta}}_i^k=\bolds{\beta}_0^k+\mathbf{b}_i^k$
as the values around which the linearization of (\ref{margi}) is
carried out at iteration $k+1$, where $\mathbf{b}_i^k$ is the
estimate of the random effect for the $i$th individual at the $k$th
iteration. This one-step algorithm is labeled LME.

This section has considered the one-sample problem, where all the
individuals share the same fixed effect vector
$\bolds{\beta}_0$. Section~\ref{section-7} considers a
two-sample model
where the direction of the subtalar axis is allowed to vary between
samples. The directional mixed effects model for this problem is
easily constructed. The local linear mixed effects model at step 2
of the Lindstrom and Bates (\citeyear{Lindstrom1990}) algorithm has a $7 \times1$ vector of
fixed regression parameters featuring 5 entries for the mean angles
in the first sample and two parameters for the between group
differences of the two subtalar angles. The two algorithms proposed
in this section can be used to estimate the parameters of this
enlarged model.

\section{A simulation study}
\label{section-6}

This section reports the results of a Monte Carlo experiment to
investigate the sampling properties of the estimators of
$\bolds{\beta}_0$ and $\bolds{\Sigma}_0$ obtained by
maximizing (\ref{margi}) with the two versions of the Lindstrom and
Bates (\citeyear{Lindstrom1990}) algorithm. First, the method used to simulate the data
is reviewed, then some results will be presented. In the simulations
the calculations are carried out with angles expressed in radians;
for convenience the results are presented in degrees.

Simulations were carried out for the one-sample model only. Values
of $M=30,50,100$ and $n=50,100,200$ were considered. The simulation
used 500 Monte Carlo samples. The following parameter values were
used:
%
\begin{eqnarray}\label{inman}
\bolds{\beta}_0&=&(8,-6,42,23,17){^\top},\qquad
\bolds{\Sigma}_0=\operatorname{diag}(7,4,9,11,11)^2,\nonumber
\\[-8pt]
\\[-8pt]
1/(2\kappa)&=&1.\nonumber
\end{eqnarray}

The mean and standard deviations for the first four anatomical
angles are as given by Inman (\citeyear{Inman1976}). The residual standard error
$1/\sqrt{2\kappa}$ of one degree was similar to estimates found in
the numerical examples of Rivest, Baillargeon and Pierrynowski (\citeyear{Rivest2008}).

For each individual, the five anatomical angles were first simulated
from a $\mathcal{N}_5 (\bolds{\beta}_0, \bolds{\Sigma
}_0)$ and the
rotation matrices $\mathbf{A}(t_1,t_2)$ and
$\mathbf{B}(s_1,s_2)$ were evaluated. To construct the predicted
values $\Psi_{ij}$ given in (\ref{psi}), angles $(\alpha_{ij},
\phi_{ij})$ obtained by fitting the one subject model to some real
data were used. The average values for $(\alpha_{ij},\phi_{ij})$
were $(38,14)$ with standard deviations of $(12,10.5)$. Thus, the
motion about the $\mathit{st}$ axis has a smaller range than that about the
$\mathit{tt}$ axis. The rotation errors $\mathbf{E}_{ij}$ were generated
from $\mathbf{z}=(z_1,z_2,z_3){^\top}$ three independent
$\mathcal{N}(0,0.017^2)$ random variables (a standard deviation of 0.017
radian is 1 when expressed in degrees). Its rotation axis was set
to $\mathbf{z}/\|\mathbf{z}\|$, while its rotation
angle was equal to $\|\mathbf{z}\|$. To understand the numerical
challenges associated to the maximization of the likelihood for the
model of Section~\ref{section-3}, it is convenient to evaluate the
vector of
partial derivatives $\mathbf{X}_{ij}$ at
$\bolds{\beta}=\bolds{\beta}_0$ in an error free model.
One gets $\mathbf{X}_{ij} = (0.01 \cos(\alpha_{ij}) - 0.99
\sin
(\alpha_{ij})$, $0.99   \cos(\alpha_{ij})$, $0.26  \cos(\phi_{ij}) + 0.95
  \sin(\phi_{ij})$, $-0.80   \cos(\phi_{ij}),1){^\top}$.
The matrix $\mathbf{X}_{i}$ of the vectors of partial derivatives
for one
subject has a condition number larger than 100. This
multi-collinearity affects especially $\mathit{ttdev}$, $\mathit{stdev}$ and
$\gamma_0$, three rotation angles about different $z$-axes that are
not well differentiated when the ankle exhibits a small range of
motion.

The simulations compared the two algorithms proposed by Lindstrom
and Bates (\citeyear{Lindstrom1990}), PLME and LME, as described in Section~\ref
{section-5}. The
\texttt{R}-function \texttt{lme} was used to fit a linear mixed
effects model at step 2 of the PLME algorithm. This function
provides estimates of the sampling variances for the fixed effects.
The biases of these variance estimators were also investigated in
the Monte Carlo study. The two algorithms gave almost the same
results. Only those obtained with PLME are presented.

\begin{sidewaystable}
\tablewidth=\textwidth
\caption{Bias and root mean squared error, in parenthesis,
of the estimator of $\bolds{\beta}_0$ and the
relative bias of the \texttt{lme} variance estimator, in parenthesis,
when the
fitted model assumes that $\bolds{\Sigma}_0$ is diagonal}\label{tab1}
\tabcolsep=0pt
\begin{tabular*}{\textwidth}{@{\extracolsep{\fill}}lccccccccccc@{}}
\hline
$\bolds n$& $\bolds M$& \multicolumn{2}{c}{$\bolds s_\mathbf{1}$} & \multicolumn{2}{c}{$\bolds s_\mathbf 2$}      & \multicolumn{2}{c}{$\bolds\gamma_\mathbf 0$}   & \multicolumn{2}{c}{$\bolds t_\mathbf 1$} & \multicolumn{2}{c@{}}{$\bolds t_\mathbf 2$} \\\hline
\phantom{2}50 & \phantom{2}30 & $-0.09$ & $(1.70, -2)$ & $0.67$ & $(3.10, -16)$ & $0.55$   & $(3.27, -14)$                    & \phantom{$-$}$ 0.10$ & $(1.56, -12)$ & $-0.09$ & $(1.26, -14)$ \\
\phantom{2}50 & \phantom{2}50 & $-0.17$ & $(1.34, -4)$ & $0.74$ & $(2.44, -13)$ & $0.54$   & $(2.74, -19)$                    & \phantom{$-$}$ 0.03$ & $(1.14, -2)$ & $-0.06$ & $(0.99, -12)$ \\
\phantom{2}50 & 100 & $-0.13$ & $(0.97, -9)$ & $0.85$ & $(1.91, -33)$ & $0.68$  & $(1.96, -23)$                    & \phantom{$-$}$ 0.05$ & $(0.85, -12)$ & $-0.12$ & $(0.73, -17)$ \\[3pt]
100 & \phantom{2}30 & $-0.09$ & $(1.58, 11)$ & $0.26$ & $(2.69, -10)$ & $0.22$  & $(2.92, -13)$                    & \phantom{$-$}$ 0.02$ & $(1.37, 0)$ &  \phantom{$-$}$0.04$ & $(1.09, -7)$ \\
100 & \phantom{2}50 & $-0.08$ & $(1.28, 4)$ & $0.52$ & $(2.13, -11)$ & $0.24$   & $(2.17, -4)$                     & \phantom{$-$}$ 0.06$ & $(1.06, 2)$ & $-0.03$ & $(0.79, 0)$ \\
100 & 100 & $-0.07$ & $(0.90, 4)$ & $0.42$ & $(1.52, -14)$ & $0.30$  &$(1.63, -17)$                     & $-0.09$ & $(0.79, -8)$ & $-0.02$ & $(0.58, -9)$ \\ [3pt]
200 & \phantom{2}30 & $-0.01$ & $(1.70, -5)$ & $0.16$ & $(2.34, -3)$ & $0.06$   &$(2.50, -9)$                      & \phantom{$-$}$ 0.06$ & $(1.38, -7)$ & \phantom{$-$}$ 0.05$ & $(0.86, 9)$ \\
200 & \phantom{2}50 & $-0.09$ & $(1.32, -5)$ & $0.39$ & $(1.87, -6)$ & $0.27$   &$(1.96, -8)$                      & $-0.02$ & $(1.03, 0)$ & \phantom{$-$}$ 0.00$ & $(0.65, 13)$\\
200 & 100 & $-0.06$ & $(0.89, 4)$ & $0.21$ & $(1.34, -9)$ & $0.06$   &$(1.44, -15)$                     & $-0.01$ & $(0.75,-5)$ & \phantom{$-$}$ 0.05$ & $(0.53, -13)$ \\
\hline
\end{tabular*}\vspace*{12pt}
%
%
%
\centering
\caption{Biases and root mean squared errors, in parenthesis, of the
estimators of the standard deviations, $\operatorname{diag} (
\bolds{\Sigma}_0^{1/2} )$,\break  when the fitted model assumes that
$\bolds{\Sigma_0}$ is diagonal}\label{tab2}
\tabcolsep=0pt
\begin{tabular*}{395pt}{@{\extracolsep{\fill}}lcccccc@{}}
\hline
$\bolds n$& $\bolds M$& $\bolds{\sigma_{t_1}}$ & $\bolds{\sigma_{t_2}}$ & $\bolds{\sigma_{s_1}}$ & $\bolds{\sigma_{s_2}}$ &
$\bolds{\sigma_{\gamma_0}}$\\
\hline
\phantom{2}50 & \phantom{1}30 & $-0.05$ $(1.10)$ & $-0.28$ $(1.28)$ & $-0.19$ $(1.25)$ &
$-0.04$ $(2.34)$
& $-0.46$ $(2.50)$\\
\phantom{2}50 & \phantom{1}50 & $-0.06$ $(0.78)$ & $-0.16$ $(0.89)$& $-0.12$ $(0.99)$ &
$-0.02$ $(1.74)$ &
$-0.18$ $(1.94)$\\
\phantom{2}50 & 100 & \phantom{$-$}$0.00$ $(0.53)$ & $-0.07$ $(0.63)$ & $-0.05$ $(0.66)$ &
\phantom{$-$}$0.03$ $(1.27)$ &
$-0.10$ $(1.26)$\\[3pt]
100 & \phantom{1}30 & $-0.11$ $(0.98)$ & $-0.11$ $(0.97)$ & $-0.07$ $(1.21)$ &
$-0.14$ $(2.09)$
& $-0.16$ $(2.27$)\\
100 & \phantom{1}50 & \phantom{$-$}$0.00$ $(0.75)$ & $-0.01$ $(0.67)$ & $-0.06$ $(0.92)$ &
\phantom{$-$}$0.13$ $(1.45)$ & $-0.02$ $(1.66)$ \\
100 & 100 & $-0.04$ $(0.54)$ & $-0.03$ $(0.49)$ & $-0.05$ $(0.70)$ & $-0.02$
$(1.12)$ & $-0.15$ $(1.20)$ \\[3pt]
200 & \phantom{1}30 & $-0.09$ $(0.96)$ & $-0.06$ $(0.76)$ & $-0.09$ $(1.14)$ &
$-0.18$ $(1.82)$ & $-0.29$ $(1.95)$ \\
200 & \phantom{1}50 & $-0.05$ $(0.75)$ & $-0.06$ $(0.58)$ & $-0.07$ $(0.92)$ &
$-0.02$ $(1.40)$ & $-0.12$ $(1.51)$\\
200 & 100 & $-0.04$ $(0.54)$ & $-0.06$ $(0.41)$ & $0$ $(0.68)$ & \phantom{$-$}$0.03$
$(0.97)$ &
$-0.05$ $(1.07)$ \\
\hline
\end{tabular*}
\end{sidewaystable}

Tables~\ref{tab1} and~\ref{tab2} report findings where the fitted model has a
diagonal $\bolds{\Sigma}_0$. In Table~\ref{tab1}, the estimates for
$t_1$, $t_2$ and $s_1$ have small biases, not significantly
different from 0, and small root mean squared errors. The estimates
for $s_2$ and $\gamma_0$ are less precise. This is caused by the
multi-collinearity problem mentioned above. The \texttt{lme}
variance estimates underestimate the true variances; this
underestimation is more severe for the angles $s_2$, $t_2$ and
$\gamma_0$. Increasing $n$, the number of data points by subject
reduces this bias. Table~\ref{tab2} is concerned with the estimation of the
standard deviations. It shows small negative biases for all the
variances. The
parameters $\sigma_{s_2}$ and $\sigma_{\gamma_0}$ have the largest
root mean squared errors. Still, Tables~\ref{tab1} and~\ref{tab2}
show that the directional mixed effects model gives reliable
estimates when the random effects are assumed to be independent.

It is likely for the anatomical angles to be correlated so that the
true value of $\bolds{\Sigma}_0$ might not be diagonal. Inman (\citeyear{Inman1976})
 did not consider this question; he did not report the
correlations between anatomical angles. One could investigate this
problem by fitting a model with an unstructured
$\bolds{\Sigma}_0$ with 15 parameters (that is the variances of
the 5 angles and the 10 covariances between pairs of angles).
Unfortunately, the results obtained with such a model are not
reliable. In simulations, not reported here, biased estimates of
the off-diagonal elements of $\bolds{\Sigma}_0$ were obtained,
especially for the covariances involving $t_2$, $s_2$ and
$\gamma_0$. Apparently, the nonlinear mixed effects model cannot distinguish
a true correlation between random effects in the population
from a correlation caused by an ill-conditioned likelihood for the
estimation of the random effects. Additional investigations of this
problem could consider models where $\bolds{\Sigma}_0$ has a
small number of nonnull covariances.

\section{Numerical examples}
\label{section-7}

This section presents the analysis of two data sets collected in the
Human Movement Laboratory of the School of Rehabilitation Science at
McMaster University, using an OptoTrak camera system at a frequency
of 50~Hz; see Rivest, Baillargeon and Pierrynowski (\citeyear{Rivest2008}) for a detailed description of the
data collection protocol. The successive rotation matrices in a data
set are not independent measurements; the residual autocorrelation
when fitting the model of Section~\ref{section-3} on the data
collected on one
subject is larger than 0.80. In order to satisfy, at least
approximately, the assumption of independence underlying the
construction of the penalized likelihood in Section~\ref{section-4},
the model was
fitted to a subsample of the data obtained by keeping one
observation out of 30. A sampling frequency of 1.67 Hz yielded
smaller residual autocorrelations, in the range $-$0.3 to 0.5; the
assumption of independence was approximately satisfied. Subsampling
the data was also used by van den Bogert, Smith
and Nigg (\citeyear{Bogert1994}) to get stable
estimates of the anatomical angles.

\subsection{\texorpdfstring{An empirical validation of the estimates
of Inman (\protect\citeyear{Inman1976})}{An empirical validation of
the estimates of Inman (1976)}}\label{sec7.1}

The mean values and the population standard deviations of the four
anatomical angles characterizing the direction of the two rotation
axes of cadaver ankles were presented by Inman (\citeyear{Inman1976}), by
manipulating unloaded cadavers feet. Inman's estimates are given in
Table~\ref{tab3}. The
model of Section~\ref{section-5} provides an in vivo method for
estimating the
same angles. In this section, we use right foot data collected on
$M=65$ volunteers with sample size $n=50$ rotation matrices. The
estimates obtained by
fitting the model of Section~\ref{section-5} to this data set are also
presented
in Table~\ref{tab3}. This model has an estimated residual standard error,
$1/\sqrt{2 \hat\kappa}$ of 0.021 radians, or 1.21 degrees.

\begin{table}
\tabcolsep=0pt
\tablewidth=294pt
\caption{The population estimates obtained by fitting the basic
model to the volunteer data set compared with the values presented
by Inman (\protect\citeyear{Inman1976})}\label{tab3}
\begin{tabular*}{294pt}{@{\extracolsep{\fill}}lcd{1.2}d{2.2}d{2.2}d{2.2}d{2.2}@{}}
 \hline
&&  \multicolumn{1}{c}{$\bolds{t_1}$}   & \multicolumn{1}{c}{$\bolds{t_2}$}   & \multicolumn{1}{c}{$\bolds{s_1}$}
& \multicolumn{1}{c}{$\bolds{s_2}$}
&\multicolumn{1}{c@{}}{$\bolds{\gamma_0}$}   \\
\hline
Data &$\hat\beta_{0j}$                  & 3.32 & -8.05 & 38.27 & 20.92 &
22.42\\
& s.e.                                  & 0.89& 1.52 & 0.89 & 2.50 & 2.37\\
& $\sqrt{\hat{\bolds{\Sigma}}_{0jj}}$   & 5.34 & 10.25 &
6.46 & 14.77 & 8.87
\\ [3pt]
Inman &$ \beta_{0j}$                    & 8 & -6 & 42 & 23 & \multicolumn{1}{c@{}}{NA}\\[1pt]
& $\sqrt{\bolds{\Sigma}_{0jj}}$         & 7 & 4 & 9 & 11 & \multicolumn{1}{c@{}}{NA}
\\ \hline
\end{tabular*}
\end{table}
%

Although two of four of Inman's values are outside of the 95\%
confidence interval, the
agreement between the two sets of estimates is reasonably good (see
Table~\ref{tab3}). The
standard deviations are
remarkably close, except possibly for the angle $t_2 = \mathit{ttdev}$. Overall,
Inman's estimates and the one derived using the directional mixed
effects model
with unloaded ankle motion data are similar.

To investigate the Bayesian model of Section~\ref{section-4}, two sets of
estimates of the anatomical angles of the right ankle of the 65
experimental subjects were calculated. The first set used $n=50$
observations per subject and the Bayesian penalty using
(\ref{inman}) as the parameters for the prior distribution. The
second set was obtained by fitting the unpenalized model of Rivest, Baillargeon and Pierrynowski (\citeyear{Rivest2008})
 to the complete data sets of $n=1500$ frames per
subject. For the two sets of estimates, the mean values for the
five angles were similar to the estimate $\hat{
\bolds{\beta}}_{0}$ in Table~\ref{tab3}. There were important
differences in the between subject standard deviations: for the
first set they were $(5.15, 6.62, 6.61, 12.84, 8.25),$ respectively,
for $(t_1,t_2,s_1,s_2, \gamma_0)$, while for the second it was
$(14.72, 18.70, 15.94, 36.56, 28.54)$. Thus, the $n=1500$ estimates
are much more variable than their penalized alternatives. The added
variability is caused by the numerical problems in maximizing the
likelihood of the unpenalized model. Similar numerical problems were
encountered by van den Bogert, Smith
and Nigg (\citeyear{Bogert1994}) when they fitted the
ankle model using an ad hoc loss function. They proposed setting
$s_2=0$ to get stable estimates. The penalty of the
Bayesian model allows to get meaningful values for the five angles.

\begin{figure}

\includegraphics{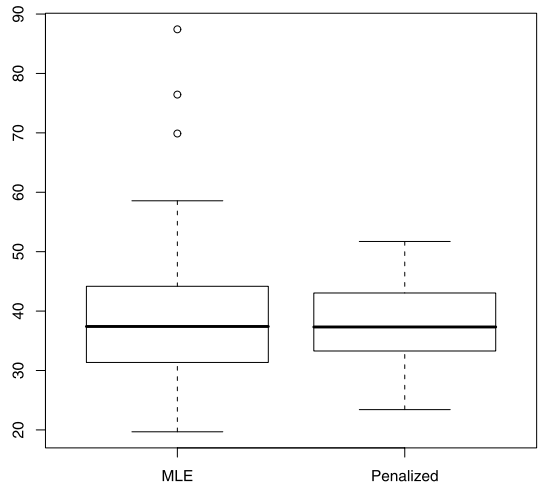}

\caption{Boxplots for two sets of estimates for
$\mathit{stinc}$ expressed in degrees.}\label{stinc}\vspace*{4pt}
\end{figure}

Several studies have found that $s_1=\mathit{stinc}$ is the only angle that
can be reliably estimated using either the model of Rivest, Baillargeon and Pierrynowski (\citeyear{Rivest2008})
 or the approach of van den Bogert, Smith
and Nigg (\citeyear{Bogert1994}). Figure~\ref{stinc} gives the
boxplots of the two sets of estimates for $s_1$; an outlier with a
negative estimate for $s_1$ has been removed. The penalty
protects against values of $s_1 = \mathit{stinc}$ larger than 70 degrees that are
not possible anatomically.

\subsection{Data analysis for the location of lower extremity injuries}\label{sec7.2}

Pierrynowski et al. (\citeyear{Pierrynowski2003}) collected ankle motion data from 31
participants who had experienced either knee ($n_1=15$) or foot
($n_2=16$) injuries during weight-bearing activities. The
experiment's goal was to determine whether there were significant
differences in the orientation of the subtalar axis between the two
groups. This section uses the data collected on the right ankle; the
individual sample size is $n_i=50$ for each subject.

\begin{table}[b]
\tabcolsep=0pt
\caption{The estimates obtained by fitting three models to the lower
extremity injury data}\label{tab4}
\begin{tabular*}{288pt}{@{\extracolsep{\fill}}lcccccccc@{}}
\hline
\textbf{Model} & & $\bolds{t_1}$   & $\bolds{t_2}$   & $\bolds{s_1}$   & $\bolds{ds_1}$   & $\bolds{s_2}$   & $\bolds{ds_2}$
&$\bolds{\gamma_0}$   \\
 \hline
1 &$\hat\beta_j$& $-4.14$& 7.09 & 45.93& $-7.41$ & \phantom{$-$}6.69& $-2.89$&
\phantom{1}$-7.33$
\\
1 & s.e. & \phantom{$-$}1.49 & 1.99 & \phantom{1}1.05& \phantom{$-$}1.42 & \phantom{$-$}3.07 & \phantom{$-$}2.56& \phantom{$>$1}3.22\\
1 & $\sqrt{\hat{\bolds{\Sigma}}_{0jj}}$ & \phantom{$-$}5.86 &  6.15& \phantom{1}3.68 &
NA & \phantom{$-$}4.68 & NA &
${<}10^{-2}$
\\ [3pt]
2 &$\hat\beta_j$& $-5.46$ & 7.62 & 45.78& $-7.45$ & \phantom{$-$}3.44& NA &
\phantom{1}$-8.89$\\
2 & s.e. & \phantom{$-$}1.41 & 1.78 & \phantom{1}0.99& \phantom{$-$}1.39 & \phantom{$-$}2.25 & NA & \phantom{$>$1}2.54\\
2 & $\sqrt{\hat{\bolds{\Sigma}}_{0jj}}$ & \phantom{$-$}5.69 & 5.40& \phantom{1}3.58 &
NA & \phantom{$-$}3.61 & NA &
${<} 10^{-2}$ \\  [3pt]
3 &$\hat\beta_j$& $-4.71$ & 5.94 & 43.01& NA & $-0.04$& NA & \phantom{1}$-9.75$\\
3 & s.e. & \phantom{$-$}1.49 & 1.97 & \phantom{1}0.96& NA & \phantom{$-$}2.69 & NA & \phantom{$>$1}2.67\\
3 & $\sqrt{\hat{\bolds{\Sigma}}_{0jj}}$ & \phantom{$-$}5.89 & 7.63& \phantom{1}4.88 &
NA & \phantom{$-$}6.35 & NA & \phantom{$>$1}4.76
\\
\hline
\end{tabular*}\vspace*{4pt}
\end{table}

Table~\ref{tab4} presents the estimates obtained by fitting three models to
this data; all the models have five independent random effects, one
for each component of $\bolds{\beta}$. The first one has 7
fixed parameters; in addition to $\bolds{\beta}$, it features
parameters $ds_1$ and $ds_2$ for the foot-knee differences of the
two subtalar angles. Model 1\ postulates that the mean $\mathit{stinc}$ and
$\mathit{stdev}$ values are $(s_1,s_2)$ and $(s_1+ds_1,s_2+ds_2)$ in the knee
and the foot group respectively; the variances of these two angles
are the same for both groups. Models 2 and 3 are derived from model
1 by setting $ds_2=0$ and $ds_1=ds_2=0$, respectively. Under model 3,
the mean values of the five angles of the ankle model are the same
in both groups. The LME and the PLME algorithms gave slightly
different numerical results; the PLME estimates are presented as a
larger maximum for the likelihood $L_{1p}$ was obtained with this
algorithm. The estimated residual standard error $1/\sqrt{2 \hat
\kappa}$ was 0.024 radians, or 1.36 degrees for the three models.

In model 1, the $z$-statistic for testing the null hypothesis of no
$\mathit{stdev}$ effect is $Z_{\mathrm{obs}}=-2.89/2.56=-1.13$ for a $p$-value of 0.26.
This high $p$-value suggests setting $ds_2=0$; this leads to model 2
where the test
for a $s_1$ between group difference has a $p$-value smaller than
$10^{-4}$; this is highly significant, even when accounting for the
underestimation of the standard errors highlighted in Table~\ref{tab1}. So
model 2 provides the best fit, in agreement with the findings of
Pierrynowski et al. (\citeyear{Pierrynowski2003}) who also noted the significant difference
in $\mathit{stinc}$ between the two groups. The nonlinear mixed effects
model allows to test for an $s_2 = \mathit{stdev}$ difference;
this was not possible with the unpenalized estimates since
they were numerically unstable.

\section{A reduced model}\label{sec8}

In his seminal work, Inman dealt with four angles, $\mathit{ttinc}$, $\mathit{ttdev}$,
$\mathit{stinc}$ and $\mathit{stdev}$. The fifth angle $\gamma_0$ of (\ref{psi})
does not play any role in his investigations. This section suggests
a reduced model featuring 4 anatomical angles instead of 5. It
investigates whether this model leads to better estimates of the
anatomical angles.

In the reference position the leg and the foot reference frames have
the same orientation, thus, the predicted value
$\bolds{\Psi}=\mathbf{I}$ is possible. Therefore, for some
angles~$\alpha$ and $\phi$, one has
\[
\mathbf{I}=\mathbf{A}(t_1,t_2)\mathbf{R}(\alpha
,x)\mathbf{R}(\gamma_0,z)
\mathbf{R}(\phi,y)\mathbf{B}(s_1,s_2){^\top}.
\]
This equation implies that $\gamma_0$ is the $Z$-Cardan angle of
$\mathbf{A}(t_1,t_2)^\top\mathbf{B}(s_1,s_2)$ in the $X-Z-Y$
convention. Thus, $\gamma_0$ can be expressed in terms of
$\mathbf{A}(t_1,t_2)$ and $\mathbf{B}(s_1,s_2)$ as
%
\begin{equation}\label{reduced}
\gamma_0(\mathbf{A}_1,
\mathbf{B}_2)=-\arcsin(\mathbf{A}_1^\top
\mathbf{B}_2).
\end{equation}

Fitting a reduced model where $\gamma_0$ depends of
$(s_1,s_2,t_1,t_2)$ through
(\ref{reduced}) is easily carried out in the framework of Section~\ref{section-3}.
All the previous
derivations hold with the reduced model provided that the matrix
$\mathbf{X}$ is redefined  in such\ a way that its $i$th row is
given by the $4 \times1$ vector:
%
\begin{eqnarray*}
\mathbf{X}_i &=& -\cos\{(\theta_{i}^z - \gamma_0)/2\}
 \\
&&\phantom{-}{}\times\pmatrix{
- \biggl(\frac{\sin(t_1) \tan
(t_2)}{ D_t^2}   \mathbf{A}_2+\frac{
1}{ D_t }
\mathbf{A}_3 \biggr)^\top \biggl(\frac{\mathbf{R}_i}
{ \sqrt{1-( \mathbf{A}_1{^\top} \mathbf{R}_i
\mathbf{B}_2)^2}}- \frac{\mathbf
{I}}{\cos\gamma_0} \biggr) \mathbf{B}_2 \cr
\frac{\cos
(t_1)\{1+ \tan^2(t_2)\}}{ D_t^2}
\mathbf{A}_2{^\top}\biggl (\frac{\mathbf{R}_i} {
\sqrt{1-( \mathbf{A}_1{^\top} \mathbf{R}_i
\mathbf{B}_2)^2}}- \frac{\mathbf
{I}}{\cos\gamma_0} \biggr) \mathbf{B}_2 \cr
\mathbf{A}_1{^\top}  \biggl(\frac{\mathbf
{R}_i} {
\sqrt{1-( \mathbf{A}_1{^\top} \mathbf{R}_i
\mathbf{B}_2)^2}}-
\frac{\mathbf{I}}{\cos
\gamma_0} \biggr)\biggl (\frac{\sin(s_1) \tan
(s_2)}{ D_s^2}  \mathbf{B}_1 +\frac{
1}{ D_s }\mathbf{B}_3 \biggr)\cr
- \frac{\cos
(s_1)\{1+ \tan^2(s_2)\}}{ D_s^2}
\mathbf{A}_1{^\top}  \biggl(\frac{\mathbf
{R}_i} {
\sqrt{1-( \mathbf{A}_1{^\top} \mathbf{R}_i
\mathbf{B}_2)^2}}-
\frac{\mathbf{I}}{\cos
\gamma_0} \biggr)\mathbf{B}_1
}
 .
\end{eqnarray*}

The reduced model has been applied to the data on the 65 volunteers
presented in Section~\ref{sec7.1}. It yielded poor results; the within
subject variability was larger than that with the five-parameter
model of Section~\ref{section-4}. The average values failed to reproduce Inman's
results. To investigate this failure, we calculated the differences
$\hat\gamma_0 + \arcsin(\hat{\mathbf{A}}_1^\top
\hat{\mathbf{B}}_2)$ obtained with the five-parameter model
for the 65 subjects of Section~\ref{sec7.1}. The average difference
is $-$2.30 degree (s.e.${}=0.34$); only 7 of the 65 differences have
positive values.

The assumption underlying the reduced model, that the leg and
the foot are aligned in the reference position, is not met.
The reference position is measured when the experimental
subject is standing, so that its ankle is loaded. However
, the data is collected on an unloaded ankle moving freely when
the subject is sitting. The nonnull value of $ \gamma_0 + \arcsin
({\mathbf{A}}_1^\top
{\mathbf{B}}_2)$ might be explained by a slight change
of the relative orientation of the two reference frames when the
ankle goes from a loaded to an unloaded position. Thus, the
reduced model is not suitable for the data
analyzed in this work. It should, however, be considered when
investigating the rotation axes of a loaded ankle. The value of $
\gamma_0 + \arcsin({\mathbf{A}}_1^\top
{\mathbf{B}}_2)$ may quantify rearfoot flexibility which is of
interest to foot
care professionals [Mansour et al. (\citeyear{Mansour2007})].

\section{Discussion}
\label{section-9}

This work presented a solution to the estimation of the directions
of the two rotation axes of the ankle. The key element is the
estimation criterion given by a penalized likelihood. This
penalized likelihood is associated to a Bayesian model for the ankle
joint and to a nonlinear mixed effects directional model that
allows estimation of the between ankle variability of the rotation
axes within a population. Simulations have shown that the population
means and the population variances can be estimated in a reliable
way. When used on a data set collected on a sample of volunteers,
the nonlinear mixed effects directional model produced mean and
variance estimates that were similar to those presented by Inman (\citeyear{Inman1976}).
The good match with Inman's clinically accepted findings
(see Table~\ref{tab3}) provides empirical evidence that a two-axis (revolute)
mechanistic model of the ankle [see equation~\eqref{psi}] is indeed
appropriate for the ankle. Section~\ref{section-7} shows that the proposed
nonlinear mixed effects directional model can be extended to
compare the ankle's axes in several populations. The Bayesian model
of Section~\ref{section-4} might very well solve the problem of
estimating in vivo
the location of the ankle's rotation axes.

Future work of this model includes a detailed investigation of the
within subject stability of the Bayesian estimates of
Section~\ref{section-4}, using both right and left foot data. The
estimation of the translation parameters of the van den Bogert, Smith
and Nigg (\citeyear{Bogert1994}) ankle's model will also be investigated and the application
of the reduced model to data collected on a loaded ankle will also
be considered.

\section*{Acknowledgments}
We are grateful to a referee for his perceptive comments who motivated
Section~\ref{sec8} and to
Sophie Baillargeon for carrying out the reduced model analysis.

\printaddresses

\end{document}